\documentclass[a4paper,10pt,twoside]{cpc-hepnp}
\usepackage{CJK,upgreek,fancyhdr}
\usepackage{multicol}
\usepackage{graphicx}
\usepackage{booktabs}
\usepackage{amssymb,bm,mathrsfs,bbm,amscd}
\usepackage[tbtags]{amsmath}
\usepackage{lastpage}
\usepackage{longtable}
\usepackage{changes}
\usepackage{lineno,hyperref}
\usepackage{threeparttable} 
\usepackage{multirow}
\usepackage{subfigure} 
\usepackage{ulem}
\usepackage{color}
\usepackage{amsmath}
\begin{document}
\begin{CJK*}{GB}{gbsn}

\fancyhead[c]{\small Chinese Physics C~~~Vol. 45, No.10 (2021) 104102}
\fancyfoot[C]{\small 010201-\thepage}


\title{Systematic study of two-proton radioactivity with a screened electrostatic barrier
\thanks{We would like to thank X. -D. Sun, J. -G. Deng, J. -H. Cheng, and J. -L. Chen for their useful discussions and input. This work is supported in part by the National Natural Science Foundation of China (Grants No. 11205083, No.11505100 and No. 11705055), the Construct Program of the Key Discipline in Hunan Province, the Research Foundation of Education Bureau of Hunan Province, China (Grants No.18A237), the Natural Science Foundation of Hunan Province, China (Grants No. 2018JJ3324), the Innovation Group of Nuclear and Particle Physics in USC, the Shandong Province Natural Science Foundation, China (Grant No. ZR2015AQ007), the National Innovation Training Foundation of China (Grant No.201910555161), and the Opening Project of Cooperative Innovation Center for Nuclear Fuel Cycle Technology and Equipment, University of South China (Grant No. 2019KFZ10).}}
\author{%
      You-Tian Zou  (×ÞÓÐÌð)$^{1}$ 
\quad Xiao Pan  (ÅËÏö)$^{1}$ 
\quad Xiao-Hua Li (ÀîС»ª) $^{1,3,4;1)}$\email{lixiaohuaphysics@126.com}\\
\quad Hong-Ming Liu (ÁõºêÃú) $^{1;2)}$\email{liuhongming13@126.com}
\quad Xi-Jun Wu(Îâϲ¾ü)$^{2;3)}$\email{wuxijun1980@yahoo.cn}
\quad Biao He (ºÎ±ë)$^{5}$
}
\maketitle 
\address{%
$^1$ School of Nuclear Science and Technology, University of South China, Hengyang 421001, China\\
$^2$School of Math and Physics, University of South China, Hengyang 421001, China \\
$^3$ Cooperative Innovation Center for Nuclear Fuel Cycle Technology $\&$ Equipment, University of South China, Hengyang 421001, China\\
$^4$ Key Laboratory of Low Dimensional Quantum Structures and Quantum Control, Hunan Normal University, Changsha 410081, China\\
$^5$College of Physics and Electronics, Central South University, Changsha 410083, China \\
}

\begin{abstract}
In this study, a phenomenological model is proposed based on Wentzel-Kramers-Brillouin (WKB) theory and applied to investigate the two-proton ($2p$) radioactive half-lives of nuclei near or beyond the proton drip line. The total diproton-daughter nucleus interaction potential is composed of the Hulthen-type electrostatic term and the centrifugal term. The calculated $2p$ radioactive half-lives can accurately reproduce the existing 10 experimental datasets of five true $2p$ radioactive nuclei with $\sigma$ = 0.736. In addition, we extend this model to predict the half-lives of possible $2p$ radioactive nuclei whose $2p$ radioactivity is energetically allowed or observed but not yet quantified in NUBASE2016. The predicted results are in agreement with those obtained using the Gamow-like model, generalized liquid drop model, Sreeja formula, and Liu formula.
\end{abstract}



\footnotetext[0]{\hspace*{-3mm}\raisebox{0.3ex}{$\scriptstyle\copyright$}2021
Chinese Physical Society and the Institute of High Energy Physics
of the Chinese Academy of Sciences and the Institute
of Modern Physics of the Chinese Academy of Sciences and IOP Publishing Ltd}%

\begin{multicols}{2}

\section{Introduction}
\label{section 1}
The stability of nuclei in the ground state depends on a complicated balance between the number of protons and neutrons. In the vicinity of  the proton drop line, where this balance is strongly disturbed, the nuclear force no longer binds the additional nucleon. However, in the case of proton-rich nuclei, the Coulomb potential barrier resulting from the charged protons can keep one (for odd-Z nuclei) or two (for even-Z nuclei) additional protons inside the nucleus for a finite time \cite{001}. These protons will eventually be emitted through the Coulomb barrier, leading to proton or two-proton ($2p$) radioactivity. These two decay modes were predicted by Goldansky $\emph{et\ al}$ using the isobaric invariance principle and the isotopic invariance principle in the 1960s \cite{1,2,3}. Proton radioactivity was initially discovered by Jackson $\emph{et\ al}$ from the isomeric state of $^{53}\rm{Co}$ in 1970 \cite{301,302}. In terms of $2p$ radioactivity, in theory, it can be divided into two cases depending on the single-proton emission released energy. One is not true $2p$ radioactivity where $Q_{2p}$$>$ 0 and $Q_{p}$$>$ 0 ($Q_{2p}$ and $Q_p$ are the released energies of $2p$ radioactivity and the single-proton emission, respectively), meaning the two emitted protons actually sequentially decay. The other is true $2p$ radioactivity where $Q_{2p}$$>$ 0 and $Q_{p}$$<$ 0 \cite{4}, meaning the two protons are emitted simultaneously. Not true $2p$ radioactivity was initially observed by a series of extremely short-lived ground-state $2p$ radioactivity emitters before 2002, such as $^6\rm{Be}$ \cite{5}, $^{12}\rm{O}$ \cite{6} and $^{16}\rm{Ne}$ \cite{6}. With the development of experimental facilities and radioactive beams, true $2p$ radioactivity was first reported in 2002 via the decay of $^{45}\rm{Fe}$ in experiments performed at GANIL \cite{8} and GSI \cite{9}, respectively. Later, true $2p$ radioactivity of $^{19}\rm{Mg}$, $^{48}\rm{Ni}$, $^{54}\rm{Zn}$ and $^{67}\rm{Kr}$ was also observed in different experiments \cite{10,11,12,13}.
\par For the true $2p$ radioactivity process, the protons can emit from two kinds of states of the parent nucleus. One state involves an isotropic emission with no angular correlation, while the other has a strong  correlation occurring as a $^2\rm{He}$-like emission from the parent nuclei. Based on these two descriptions and Wentzel-Kramers-Brillouin (WKB) theory, a number of theoretical models and/or empirical formulas were proposed for investigation of $2p$ radioactivity, such as the direct decay model \cite{14,15,16}, the diproton model \cite{15,18,19,24,25,255,26}, the three-body model \cite{20,21,22,23}, the empirical formulas of Sreeja \cite{31} and Liu \cite{32} and others \cite{27,28,29,30}. These theoretical approaches have partially improved our understanding of the $2p$ radioactivity phenomenon. However, the calculations in these theoretical approaches do not take into account the electrostatic screening effect caused by the superposition of the involved charges, the magnetic field generated by the movement of the two emitted protons or the inhomogeneous charge distribution of the nucleus \cite{366,35,36}. This effect was first considered by Hulthen $\emph{et\ al}$ in 1942 \cite{33}. In their study, they proposed an analytic form of electrostatic interaction to describe this effect named as Hulthen potential. In recent years, this potential has been extensively employed to study the half-lives of $\alpha$ decay and proton radioactivity\cite{35,36,361,362}. The calculated results reproduce the experimental data well. The $2p$ radioactivity process shares the same theory, i.e., barrier penetration as $\alpha$ decay and proton radioactivity \cite{363,364,3641,3642,3643,3644,3645,3646,3647,3648}. Whether the Hulthen potential can be extended to study $2p$ radioactive half-lives is an interesting topic. In this study, based on the WKB theory and employing the Hulthen potential as a replacement for Coulomb potential to consider the electrostatic screening effect, we systematically investigate the half-lives of $2p$ radioactive nuclei near or beyond the proton drip line.
\par The article is arranged as follows. In the next section, the theoretical framework is briefly presented. The detailed calculations and discussion are presented in Section 3. Finally, a summary is given in Section 4.

\section{Theoretical framework}
\label{section 2}

The $2p$ radioactive half-life $T_{1/2}$ can be calculated using the decay constant ${\lambda}$ and expressed as
\begin{equation}
T_{1/2} = \frac{ ln2}{\lambda}
\label{eq1}
\end{equation}

with
\begin{equation}
\lambda = S_{2p} \nu \emph{P}.
\label{eqq1}
\end{equation}

Here, $S_{2p}$ = $G_0^2[A/(A-2) ]^{2n}\upvarrho^2$ denotes the spectroscopic factor of $2p$ radioactivity estimated using the cluster overlap approximation \cite{19} with $G_0^2$ = $(2n)!/[2^{2n}(n!)^2]$ \cite{38}, where n $\approx$ $(3Z)^{1/3}-1$ is the average principal proton oscillator quantum number \cite{39}, $A$ and $Z$ denote the mass number and proton number of the parent nucleus, respectively, $\upvarrho^2$= $0.015$ denotes the proton overlap function, which is determined by a $\upchi^2$ optimization from the experimental half-lives of $^{19}\rm{Mg}$, $^{45}\rm{Fe}$, $^{48}\rm{Ni}$ and $^{54}\rm{Zn}$ \cite{14}, and $\nu$ is the assault frequency, which can be calculated by the oscillation frequency $\omega$ and expressed as \cite{41}
\begin{equation}
\nu = \frac{\omega}{2\pi} = \frac{(2n_r + l +\frac{3}{2})\hbar}{1.2\pi \mu R_n^2} = \frac{(G+\frac{3}{2})\hbar}{1.2\pi \mu R_{00}^2} ,
\end{equation}

where $R_n$ = $\sqrt{\frac{3}{5}}R_{00}$ is the nucleus root-mean-square (rms) radius with $R_{00}$ = $1.24A^{1/3}$ $(1+ \frac{1.646}{A}- 0.191\frac{A-2Z}{A})$ \cite{411}, and $G$ = $2n_r$ + $l$ is the principal quantum number with $n_r$ and $l$ being the radial and angular momentum quantum number, respectively. For $2p$ radioactivity, we set $G$ = 4 or 5 corresponding to the $4\hbar \omega$ or $5\hbar \omega$ oscillator shells, depending on the individual two-proton emitters. The reduced mass of the decaying nuclear system is given by $\mu$ = $m_d$$m_{2p}$/($m_d$ + $m_{2p}$) $\simeq$ 938.3 $\times$ 2 $\times$ $(A-2)$/$A$ MeV/$c^2$, with $m_d$ and $m_{2p}$ being the mass of daughter nucleus and the two emitted protons, respectively. Finally, $\hbar$ is the reduced Planck constant.
 \par The quantity $P$ given in Eq.({\,\ref{eqq1}}) is the penetration probability of the two emitted protons crossing the barrier. It can be calculated by the semi-classical WKB approximation \cite{42,44} and written as
\begin{equation}
P = \rm{exp}\{-\frac{2}{\hbar} \int_{\emph{R}_\emph{t}}^{\emph{R}_\emph{out}} \sqrt{2\mu[\emph{V} (\emph{r})-Q_{2\emph{p}}]} dr \} \ ,
\label{eq4}
\end{equation}

where $V(r)$ is the diproton-daughter nucleus interaction potential, $R_{out}$ is the outer turning point of potential barrier satisfying the condition $V(R_{out})$ = $Q_{2p}$, and $R_t$ = $R_d$ + $R_{2p}$ is the distance of the touching configuration, where $R_d$ and $R_{2p}$ are the radii of daughter nucleus and the two emitted protons, respectively. They can be calculated using \cite{45}
\begin{equation}
R = 1.28A^{1/3} - 0.76 + 0.8A^{-1/3}. 
\end{equation}
\par In general, the diproton-daughter nucleus electrostatic potential is by default Coulomb type, given as
\begin{equation}
V_C = {Z_dZ_{2p}e^2}/{r},
\end{equation}

where $Z_d$ and $Z_{2p}$ are the proton numbers of the daughter nucleus and the two emitted protons, respectively, and $e^2$ is the square of the electronic elementary charge. However, in the $2p$ radioactivity process, for the superposition of the involved charges, due to the magnetic field generated by the movement of the two emitted protons and the inhomogeneous charge distribution of the nucleus, the diproton-daughter nucleus electrostatic potential behaves as a Coulomb potential at a short distance and drops exponentially at large distance, i.e., the screened electrostatic effect \cite{33}. This behavior of electrostatic potential can be described using the Hulthen type potential and is defined as
  \begin{equation}
V_H = \frac{aZ_dZ_{2p}e^2}{e^{ar}-1},
\end{equation}

 where $a$ is the screening parameter. In this framework, the total diproton-daughter nucleus interaction potential $V(r)$, shown in Fig\,\ref{fig 1}, is given by
\begin{equation}
\label{eq7}
\
V(r) = \left\{\begin{array}{ll}
-V_0,  &0 \leq r \leq R_t ,\\
V_H(r) + V_l(r),  &r \ge R_t ,\\
\end{array}\right.
\end{equation}

where $V_0$ is the depth of square well and $V_l$($r$) is the centrifugal potential. 

As $l (l + 1)$ $\to$ $(l + \frac{1}{2})^2$ is a necessary correction for one-dimensional problems \cite{455}, the centrifugal potential is chosen as the Langer modified centrifugal potential in this study. It can be expressed as
\begin{equation}
V_l(r) = \frac{\hbar^2(l+\frac{1}{2})^2}{2\mu r^2},
\end{equation}

where $\emph{l}$ is the orbital angular momentum taken away by the two emitted protons. It can be obtained by the spin and parity conservation laws.

\begin{center}
\includegraphics[width=9cm]{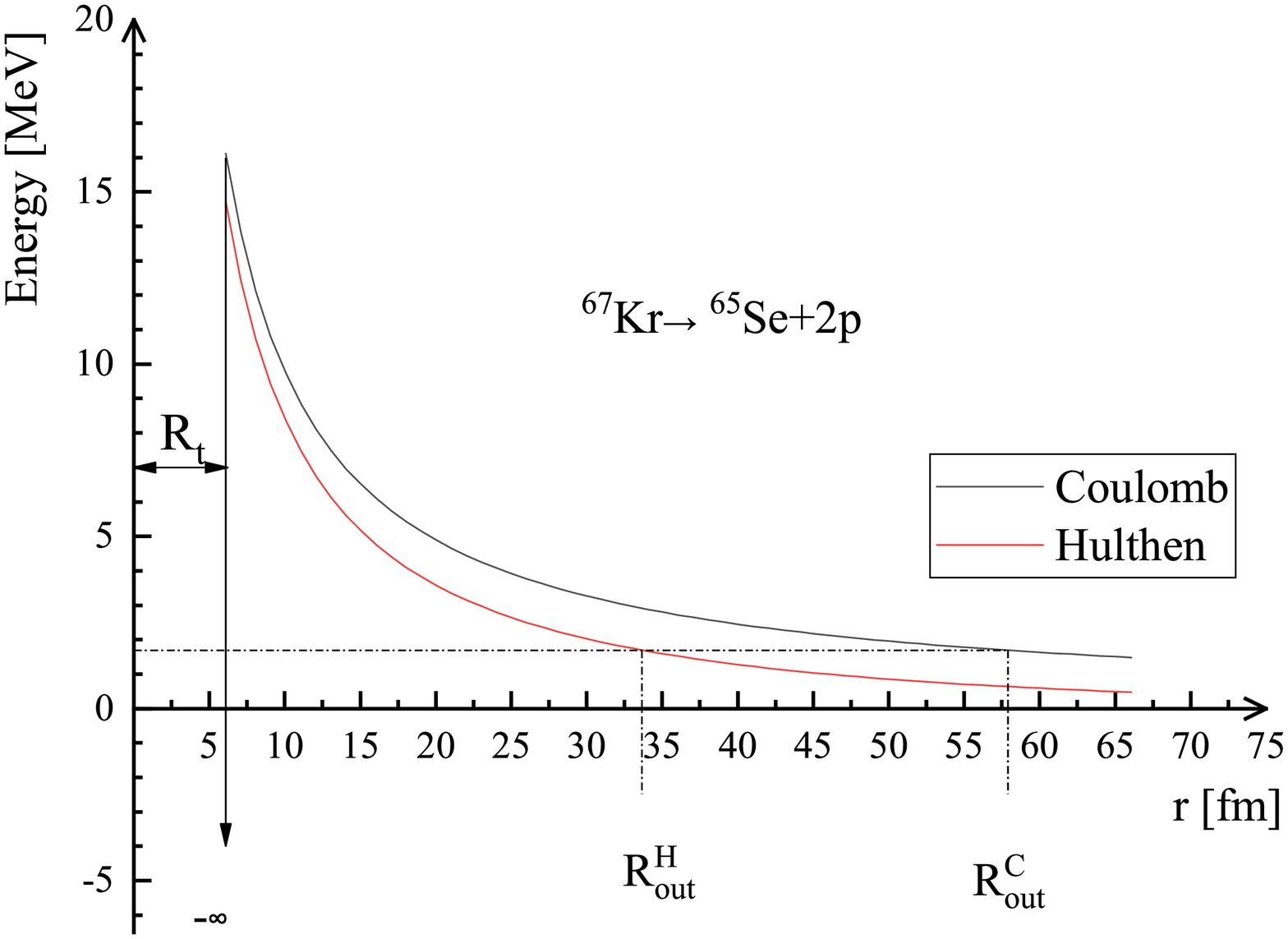}
\figcaption{\label{figure1} (color online) Schematic plot of the potential energy $V(r)$ for $^{67}\rm{Kr}$ as a function of the distance between the centers of the daughter nucleus and the two emitted protons.}
\label {fig 1}
\end{center}

Obviously, the outer turning point of potential barrier $R_{out}$ given in Eq.({\,\ref{eq4}}) is an important quantity in evaluating the half-life of $2p$ radioactivity. To obtain an analytic expression for $R_{out}$, the centrifugal term should be approximated as \cite{46}
\begin{equation}
\frac{1}{r^2} \approx \frac{a^2}{(e^{ar}-1)^2},
\end{equation}

with the value of $a$ being small. Under this assumption, the outer turning point of the potential barrier $R_{out}$ can be expressed as 
\begin{equation}
R_{out} = \frac{1}{a}\rm{ln}\left[ \frac{2V_1}{\sqrt{{V_0}^2+4V_1Q_{2p}}-V_0}+1\right]
\label{eq10}
\end{equation}
with
\begin{equation}
V_0=2ae^2Z_1, \  V_1=\frac{a^2\hbar^2(l+\frac{1}{2})^2}{2\mu} .
\end{equation}

We define the Gamow factor as $G_{se}$ = -$\log{P}$ for the outer potential barrier region. It can be given by the definite integral as \cite{35}
\begin{equation}
G_{se} = \frac{1}{a}[I_1(r)+ I_2(r)] \vert^{R_{out}}_{R_t}. 
\end{equation}

These two terms, $I_1(r)$ and $I_2(r)$, have the following expressions: \cite{35,36}
\begin{multline}
I_1(r) = -\sqrt{V_1x^2 + V_0x - Q_{2p}} \\ + \sqrt{Q_{2p}}\arcsin{\left[\frac{xV_0 - 2Q_{2p}}{x\sqrt{4Q_{2p}V_1 + {V_0}^2}}\right]} \\ - \frac{V_0}{2\sqrt{V_1}} \ln{\left[2\sqrt{\left(V_1x^2 + V_0x - Q_{2p} \right) } + V_0 +2V_1x \right]} ,
\end{multline}

\begin{multline}
I_2(r) = \sqrt{V_1y^2 + U_0y - U_1} \\ - \sqrt{U_1}\arctan{\frac{yU_0-2U_1}{2\sqrt{U_1(V_1y^2 + U_0y - U_1)}}} \\ + \frac{U_0}{2\sqrt{V_1}} \ln{\left[ 2\sqrt{V_1(V_1y^2+U_0y-U_1)}+U_0+2V_1y\right]},
\end{multline}

where the following notations are employd:
\begin{gather}
x = (e^{ar} - 1)^{-1}, \ y = 1 + (e^{ar} - 1)^{-1}, \\ \notag
U_0 = V_0 - 2V_1, \ U_1 = Q_{2p} + V_0 -V_1.       
\end{gather}

\section{Results and discussion}
In this study, we systematically investigate the $2p$ radioactive half-lives of nuclei near or beyond the proton drip line. The experimental $2p$ radioactive half-lives and corresponding released energies are taken from the corresponding references of $2p$ radioactivity. With the Hulthen potential for the electrostatic barrier being considered, this model contains one adjustable parameter, i.e., the screening parameter $a$, which is obtained by fitting the true $2p$ radioactive nuclei of $^{19}$$\rm{Mg}$, $^{45}$$\rm{Fe}$, $^{48}$$\rm{Ni}$, $^{54}$$\rm{Zn}$, and $^{67}$$\rm{Kr}$, amounting to 10 experimental datasets. The standard deviation $\sigma$, describing the difference between the experimental data and calculated data, can be expressed as
\begin{equation}
\sigma=\sqrt{\frac{1}{10}\sum_{i=1}^{10}\left[\log{\left(\frac{T^{\rm{cacl}}_{1/2_i}}{T^{\rm{expt}}_{1/2_i}}\right)}\right]^2} .
\label{eq:sgm}
\end {equation}
Here, ${T^{\rm{expt}}_{1/2_i}}$ and ${T^{\rm{cacl}}_{1/2_i}}$ represent the experimental and calculated $2p$ radioactive half-lives for the $i$-th nucleus, respectively. Through minimizing $\sigma$, the screening parameter is determined as $a$ = $2.9647\times 10^{-2} \rm{fm}^{-1}$. The value of $a$ is small, though it observably impacts the classical outer turning point $R_{out}$ obtained by eq.(\,\ref{eq10}), with the $2p$ radioactive half-life being sensitive to $R_{out}$. For intuitively displaying the screening effect, in Fig.\,\ref{fig 2}, we illustrate the difference between $R^C_{out}$ - $R^H_{out}$ against the proton number of the daughter nucleus for different $2p$ radioactivity released energies $Q_{2p}$. Here, $R^C_{out}$ and $R^H_{out}$ represent the $R_{out}$ value calculated using pure Coulomb and Hulthen potentials, respectively, while the centrifugal potential contribution is not considered. From this figure, we can clearly see that the smaller $2p$ radioactivity released energy $Q_{2p}$ and larger proton number of the daughter nucleus $Z_d$ increase the difference between $R^C_{out}$ and $R^H_{out}$.

\begin{center}
\includegraphics[width=9cm]{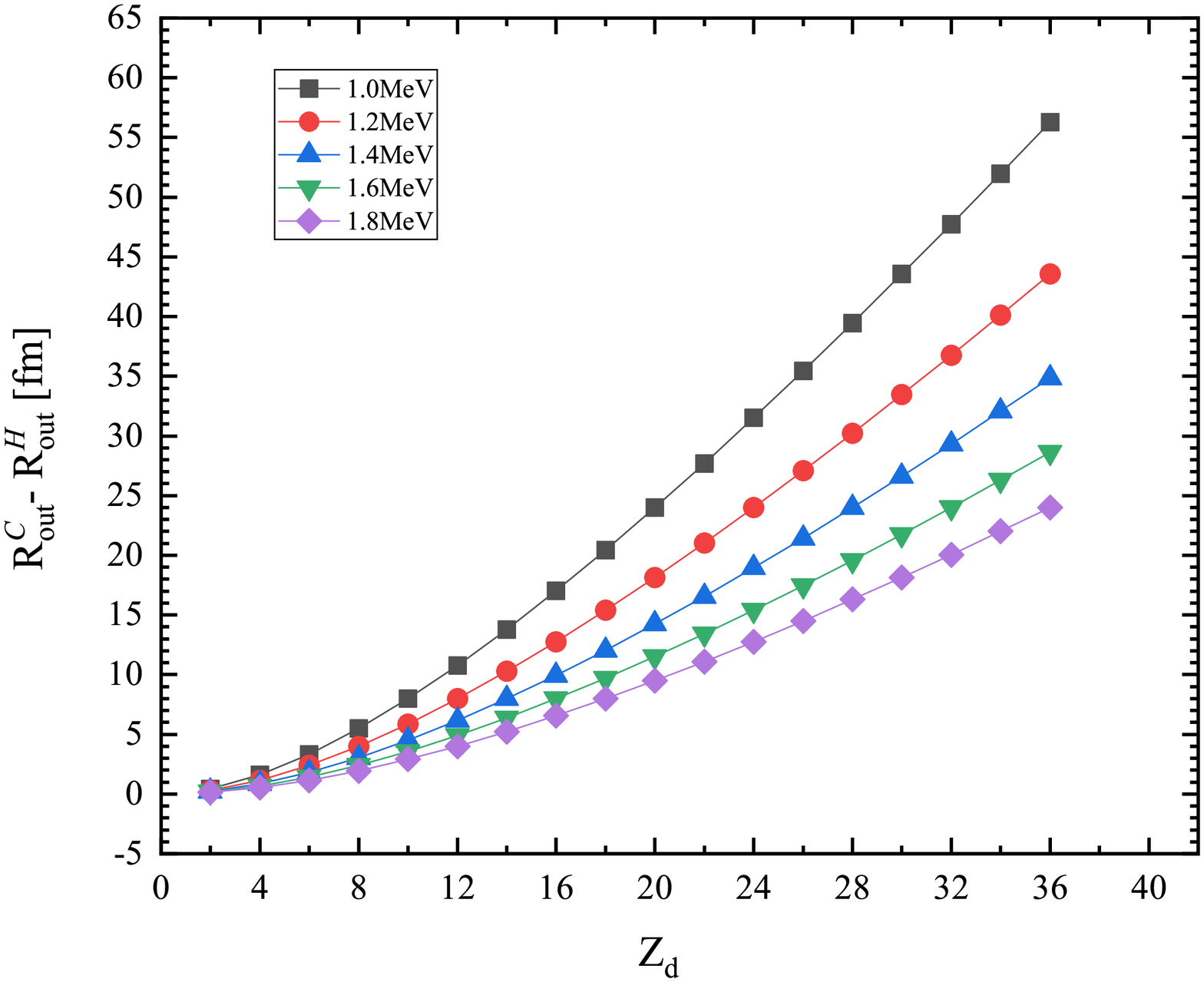}
\figcaption{\label{figure2}(color online) Difference between $R^C_{out}$ and $R^H_{out}$ obtained by V(r) = $Q_{2p}$ when only considering the Coulomb potential. For calculating-$R^C_{out}$, the Coulomb potential is taken as the potential of a uniformly charged sphere, i.e., $V_C$(r) = ${Z_dZ_{2p}e^2}/{r}$ while for $R^H_{out}$, the Coulomb potential is taken as the Hulthen potential, i.e., $V_H(r)$ = $\frac{aZ_dZ_{2p}e^2}{e^{ar}-1}$, with $a$ = $2.9647\times10^{-2} \rm{fm}^{-1}$.}
\label {fig 2}
\end{center}

For a more intuitive illustration of the relationship between the screening  effect and $Q_{2p}$, taking the true $2p$ radioactive nucleus $^{45}$\rm{Fe} as an example, we plot the difference of $R^C_{out}$ - $R^H_{out}$ against $Q_{2p}$ in Fig.\,\ref{fig 3}. From this figure, we can see an approximately linear relation between $R^C_{out}$ - $R^H_{out}$ and $Q_{2p}$, indicating a strong correlation between the screening effect and $2p$ radioactivity released energy.
\begin{center}
\includegraphics[width=9cm]{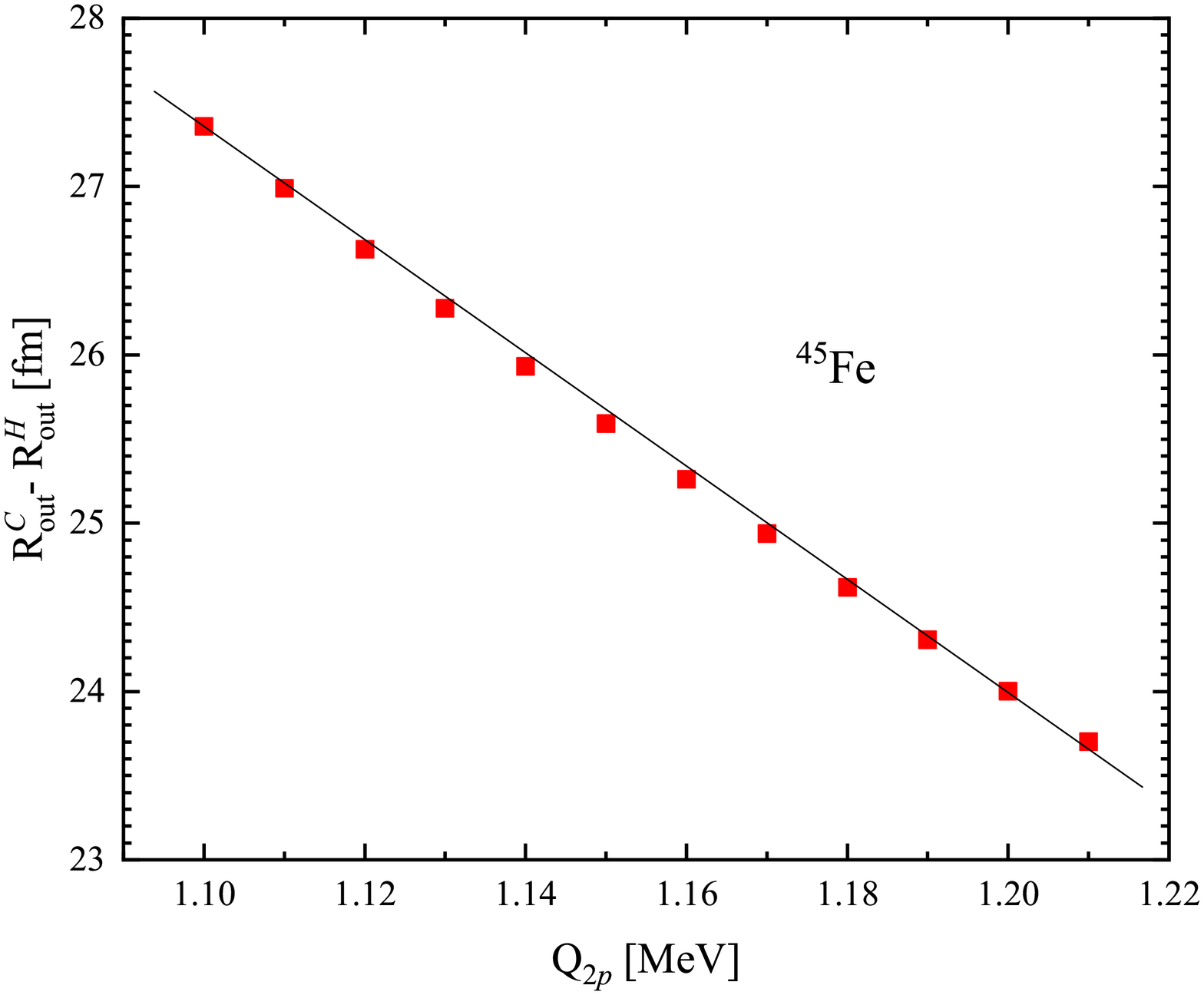}
\figcaption{\label{figure3} (color online) Difference between $R^C_{out}$ and $R^H_{out}$ plotted against the 2p radioactivity released energy $Q_{2p}$ for $^{45}$\rm{Fe}. }
\label {fig 3}
\end{center}

In the following, using this model and the value of parameter $a$, we systematically calculate the $2p$ radioactive half-lives of nuclei using the experimental data, including true and not true $2p$ radioactive nuclei. The detailed results are listed in table \,\ref{table 1}. For comparison, the Gamow-like model \cite{24}, generalized liquid drop model \cite{26}, Sreeja formula \cite{31}, and Liu formula \cite{32} are also employed. In this table, the first three columns represent the $2p$ radioactive parent nucleus, $2p$ radioactivity released energy, and the logarithmic form of experimental $2p$ radioactive half-lives (Expt), respectively. The last five columns represent the logarithmic form of theoretical $2p$ radioactive half-lives, which are calculated in this study and with the Gamow-like model (GLM) \cite{24}, generalized liquid drop model (GLDM) \cite{26}, Sreeja formula (Sreeja) \cite{31}, and Liu formula (Liu) \cite{32}, respectively. In order to intutively provide comparisons of the experimental $2p$ radioactive half-lives with the calculated values, we present the decimal logarithm deviations between the experimental $2p$ radioactive half-lives and the calculated values in Fig.\,\ref{fig 4}. From this figure, it can be seen that the decimal logarithm deviations between the experimental $2p$ radioactive half-lives and the values calculated in this study are within $\pm1$. However, for $^{54}$Zn with $Q_{2p}$ = 1.280 MeV, the decimal logarithm deviations between the experimental $2p$ radioactive half-lives and the values calculated using the theoretical approaches of the Gamow-like model, generalized liquid drop model, Sreeja formula, Liu formula, and this study approach approximately an order of magnitude. The result suggests that this experimental data may not be accurate enough for theoretical comparisons. Furthermore, in order to compare the reproducibility of our model with other theoretical models and/or formulas for $2p$ radioactive half-lives, we calculate the standard deviation $\sigma$ using Eq.\,(\ref{eq:sgm}), with the results listed in table \,\ref{table 2}. From this table, we find that the minimum $\sigma$ obtained by our study is 0.736. Compared to the best result of previous studies, the value of $\sigma$ is reduced by $\frac{0.846-0.736}{0.846}$ = $13.0\%$. It means that our study can be treated as a new and effective tool to investigate $2p$ radioactivity.

Finally, we extend this model to predict the half-lives of possible $2p$ radioactive nuclei with $Q_{2p}$ $>$ 0, which are extracted from the evaluated atomic mass table AME2016 \cite{48,49}. The predicted results with other predictions using the Gamow-like model, GLDM, and two empirical formulas of Sreeja and Liu are all listed in table \,\ref{table 3}. This table is similar to table \,\ref{table 1}, except the experimental $2p$ radioactivity half-lives denoted as ``$\rm{Expt}$'' are replaced by the orbital angular momentum denoted by $\emph{l}$ in the third column. Based on table \,\ref{table 3}, we plot the logarithmic form of the predicted $2p$ radioactive half-lives for these possible $2p$ radioactive nuclei in Fig.\,\ref{fig 5}. From this figure, we observe that our predicted results are in agreement with the results from the Gamow-like model,  generalized liquid drop model, Sreeja formulas, and Liu formulas. In addition, we plot the predicted $2p$ radioactive half-lives $\log_{10}T_{1/2}^{\rm{Pre}}$ against $( Z_d^{0.8} + l^{0.25} ) Q_{2p}^{-1/2}$, i.e., the new Geiger-Nuttall law for $2p$ radioactivity \cite{32}, in Fig.\,\ref{fig 6}. The figure depicts an approximately straight line, which indicates that our predicted results are reliable.
\end{multicols}

\begin{center}
\tabcaption{Comparisons between the experimental $2p$ radioactive half-lives and the calculated values using five different theoretical models and/or formulas. The experimental $2p$ radioactive half-lives $\log_{10}T_{1/2}^{\rm{expt}}$ and corresponding $2p$ released energies $Q_{2p}$ are extracted from the different references.}
\label {table 1}
\footnotesize
\begin{tabular}{cccccccc}
\hline \hline
\multicolumn{2}{c}{\multirow{2}{*}{}}&\multicolumn{6}{c}{$\log_{10}T_{1/2}$ (s)}\\
 \cline{3-8} 
 {Nuclei} & $Q_{2p}$ (MeV) & {Expt} &{This study} &{GLM} & {GLDM}&{Sreeja}& {Liu}\\ \hline 
 \noalign{\global\arrayrulewidth1.2pt}\noalign{\global\arrayrulewidth0.4pt} \multicolumn{8}{c}{\textbf{}}\\
$^{6}$Be	       &$1.371$\cite{5}	 &	$-20.30$\cite{5} 	  &	$-19.86$ 	&$	-19.70 	$&$	-19.37      $&$-21.95      $&$-23.81      $\\  
$^{12}$O       &$1.638$\cite{52}    &    $>-20.20$\cite{52}  &  $-17.70$  &$	-18.04	$&$	-19.71 	$&$	-18.47	$&$-20.17 	$\\
$$                   &$1.820$\cite{2} 	&     $-20.94$\cite{2}   &  $-18.03$  &$	-18.30 	$&$	-19.46 	$&$	-18.79 	$&$-20.52 	$\\
$$                 &$1.790$\cite{54} 	&     $-20.10$\cite{54}   &  $-17.98$  &$	-18.26 	$&$	-19.43 	$&$	-18.74	$&$-20.46	$\\
$$                &$1.800$\cite{55} 	&     $-20.12$\cite{55}   &  $-18.00$  &$	-18.27 	$&$	-19.44 	$&$	-18.76 	$&$-20.48 	$\\  
$^{16}$Ne  &$1.330$\cite{2} 	&     $-20.64$\cite{2}   &  $-15.47$  &$	-16.23	$&$	-16.45 	$&$-15.94	$&$-17.53 	$\\
$$                &$1.400$\cite{56} 	&     $-20.38$\cite{56}   &  $-15.71$  &$	-16.43 	$&$	-16.63 	$&$-16.16 	$&$	-17.77 	$\\  
$^{19}$Mg  &$0.750$ \cite{10}	&     $-11.40$\cite{10}   &  $-10.58$  &$	-11.46	$&$	-11.79 	$&$-10.66	$&$-12.03 	$\\ 
$^{45}$Fe    &$1.100$ \cite{9}	&     $-2.40$  \cite{9}  &   $-2.32$  &$	-2.09	$&$	-2.23	$&$-1.25 	$&$-2.21 	$\\ 
$$                 &$1.140$\cite{8} 	&     $-2.07$\cite{8}     &   $-2.67$  &$	-2.58 	$&$	-2.71	$&$-1.66 	$&$	-2.64	$\\ 
$$                 &$1.154$\cite{13} 	&     $-2.55$\cite{13}    &   $-2.78$  &$	-2.74 	$&$	-2.87 	$&$-1.80 	$&$	-2.79 	$\\
$$                 &$1.210$\cite{61} 	&     $-2.42$\cite{61}     &   $-3.24$  &$	-3.37 	$&$	-3.50 	$&$-2.34	        $&$	-3.35 	$\\  
$^{48}$Ni    &$1.290$\cite{62} 	&     $-2.52$\cite{62}     &   $-2.55$  &$	-2.59	$&$	-2.62 	$&$-1.61 	$&$-2.59 	$\\ 
$$                 &$1.350$\cite{13} 	&     $-2.08$\cite{13}     &   $-3.00$  &$	-3.21 	$&$	-3.24 	$&$-2.13 	$&$	-3.13	$\\  
$^{54}$Zn    &$1.280$\cite{63} 	&     $-2.76$\cite{63}    &   $-1.31$  &$	-0.93	$&$	-0.87 	$&$-0.10 	$&$-1.01 	$\\ 
$$           &$1.480$\cite{11} 	&     $      -2.43$\cite{11}    &   $-2.81$  &$	-3.01 	$&$	-2.95 	$&$-1.83 	$&$	-2.81 	$\\  
$^{67}$Kr    &$1.690$\cite{12} 	&     $-1.70$\cite{12}    &   $-0.95$  &$	-0.76	$&$	-1.25 	$&$0.31  	$&$-0.58 	$\\ 
\hline \hline
\end{tabular}
\end{center}

\begin{center}
\includegraphics[width=14cm]{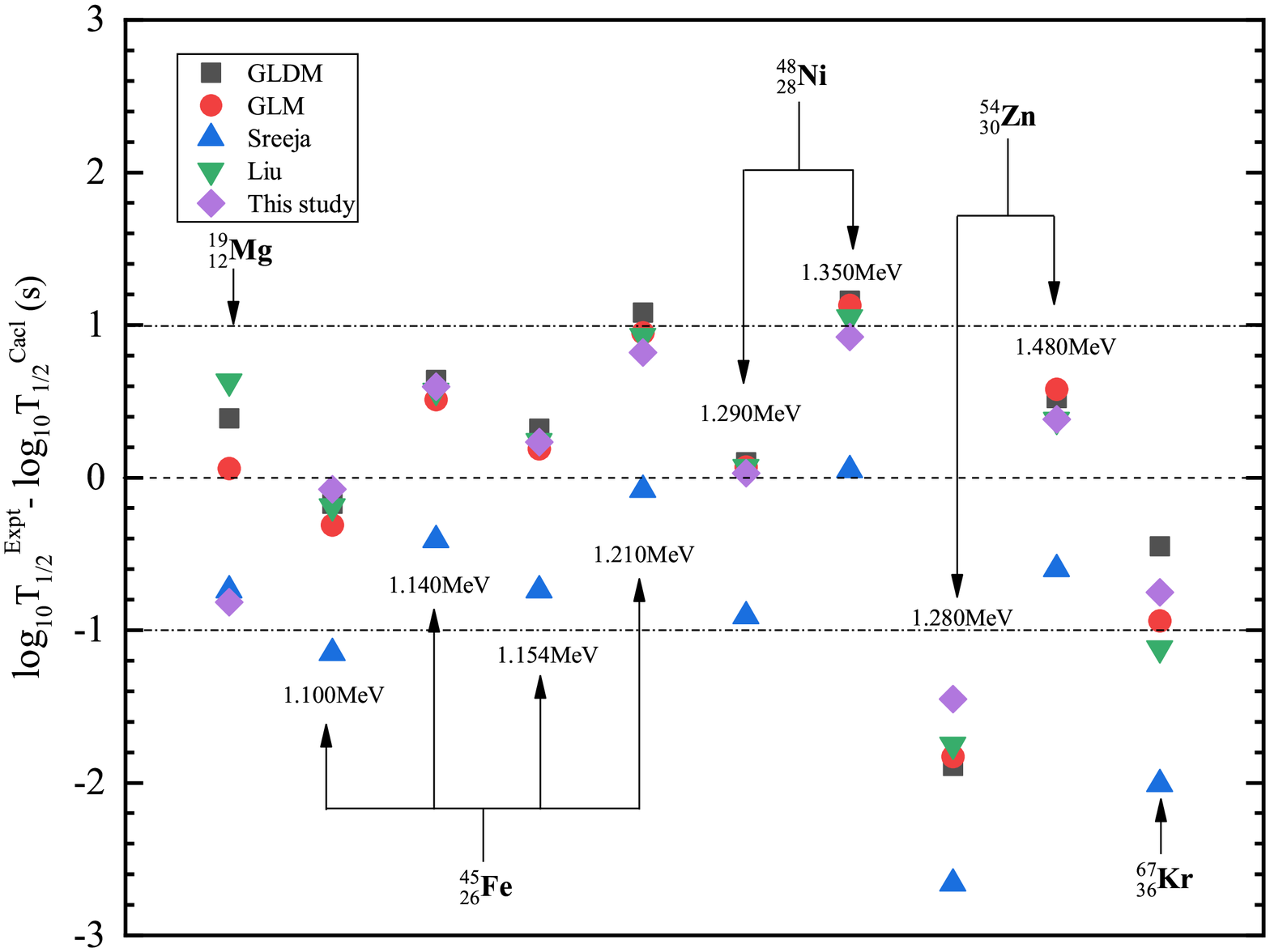}
\figcaption{\label{figure4} (color online) Deviations between the experimental $2p$ radioactive half-lives and the calculated values using different theoretical models and/or formulas for true and not true $2p$ radioactive nuclei.}
\label {fig 4}
\end{center}

\begin{center}
\tabcaption{Standard deviation $\sigma$ between the experimental data and the calculated data using different theoretical models and/or formulas for true $2p$ radioactivity.}
\label {table 2}
\footnotesize
\begin{tabular}{cccccc}
\hline
 {Model} & {This study} & {Gamow} &{GLDM} &{Sreeja} & {Liu}\\ \hline 
$\sigma$  &  $0.736$  &  $0.846$  &  $0.852$ 	&$1.221$  & $0.851$\\  
\hline 
\end{tabular}
\end{center}

\begin{center}
\tabcaption{Comparison of the predicted half-lives for possible $2p$ radioactive nuclei whose $2p$ radioactivity is energetically allowed or observed but not yet quantified in NUBASE2016 \cite{50}.}
\label {table 3}
\footnotesize
\begin{tabular}{cccccccc}
\hline \hline
\multicolumn{2}{c}{\multirow{3}{*}{}}&\multicolumn{5}{c}{$\log_{10}T_{1/2}$ (s)}\\
 \cline{4-8} 
 {Nuclei} & $Q_{2p}$ (MeV) & \emph{l}  &{This study} &{GLM} & {GLDM}&{Sreeja}& {Liu}\\ \hline 
 \noalign{\global\arrayrulewidth1.2pt}\noalign{\global\arrayrulewidth0.4pt} \multicolumn{8}{c}{\textbf{}}\\
$^{22}$Si	    &$1.283$ & 0 &	$-12.17$ 	  &	$-13.25$ 	&$	-13.30 	$&$	-12.30      $&$-13.74$   \\  
$^{26}$S	    &$1.755$ & 0 &	$-12.82$ 	  &	$-13.92$ 	&$	-14.59 	$&$	-12.71      $&$-14.16$   \\  
$^{34}$Ca	    &$1.474$ & 0 &	$ -8.99$ 	  &	$-10.10$ 	&$	-10.71 	$&$	-8.65       $&$-9.93$   \\  
$^{36}$Sc	    &$1.993$ & 0 &	$-10.79$ 	  &	$-12.00$ 	&$	  -    	$&$	-10.30      $&$-11.66$   \\  
$^{38}$Ti	    &$2.743$ & 0 &	$-12.70$ 	  &	$-13.84$ 	&$	-14.27 	$&$	-11.93      $&$-13.35$   \\  
$^{39}$Ti	    &$0.758$ & 0 &	$-1.91$ 	          &	$-0.91$ 	&$	-1.34 	$&$	-0.28       $&$-1.19$   \\  
$^{40}$V	    &$1.842$ & 0 &	$-8.97$ 	          &	$-10.15$ 	&$	  -   	$&$	-8.46       $&$-9.73$   \\  
$^{42}$Cr	    &$1.002$ & 0 &	$-2.87$ 	          &	$-2.65$ 	&$	-2.88 	$&$	-1.78       $&$-2.76$   \\  
$^{47}$Co	    &$1.042$ & 0 &	$-1.13$ 	          &	$-0.42$ 	&$	  -    	$&$	 0.21       $&$-0.69$   \\   
$^{56}$Ga	    &$2.443$ & 0 &	$-7.41$ 	          &	$-8.57$ 	&$	  -   	$&$	-6.42       $&$-7.61$   \\  
$^{58}$Ge	    &$3.732$ & 0 &	$-11.10$ 	  &	$-12.32$ 	&$	-13.10 	$&$	-9.53       $&$-10.85$   \\ 
$^{59}$Ge	    &$2.102$ & 0 &	$-5.41$ 	          &	$-6.31$ 	&$	-6.97 	$&$	-4.44       $&$-5.54$   \\  
$^{61}$As	    &$2.282$ & 0 &	$-5.78$ 	          &	$-6.76$ 	&$	  -   	$&$	-4.74       $&$-5.58$   \\  
$^{10}$N	    &$1.300$ & 1 &	$-16.76$ 	  &	$-17.36$ 	&$	  -   	$&$	-20.04      $&$-18.59$   \\  
$^{28}$Cl	    &$1.965$ & 2 &	$-11.78$ 	  &	$-13.11$ 	&$	  -   	$&$	-14.52      $&$-12.46$   \\  
$^{32}$K	    &$2.077$ & 2 &	$-11.13$ 	  &	$-12.49$ 	&$	  -   	$&$	-13.46      $&$-11.55$   \\  
$^{57}$Ga	    &$2.047$ & 2 &	$-4.94$ 	          &	$-5.91$ 	&$	  -   	$&$	-5.22       $&$-4.14$   \\  
$^{60}$As	    &$3.492$ & 4 &	$-7.88$ 	          &	$-9.40$ 	&$	  -    	$&$	-10.84      $&$-8.33$   \\  
\hline \hline
\end{tabular}
\end{center}

\begin{center}
\includegraphics[width=14cm]{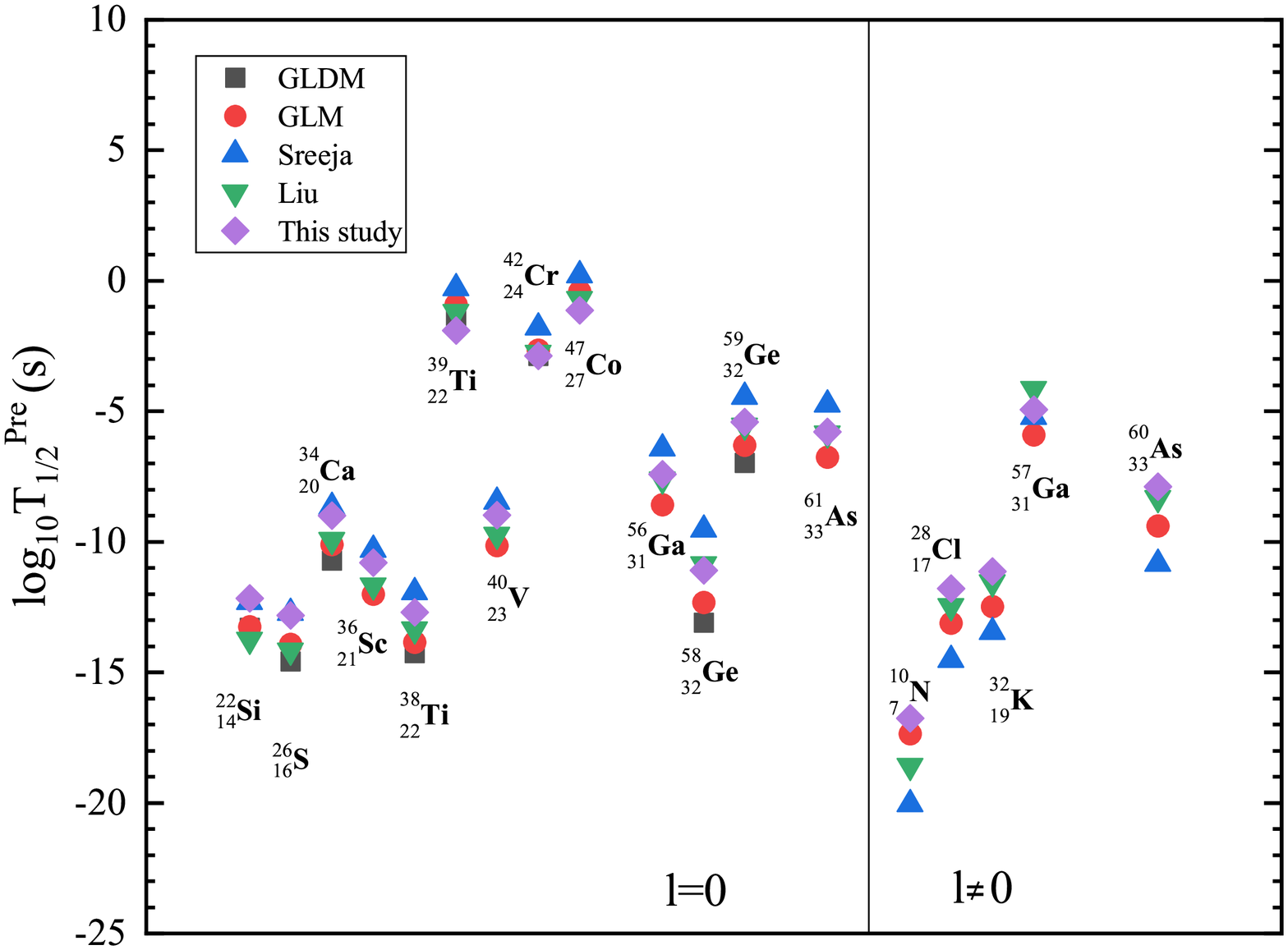}
\figcaption{\label{figure5} (color online) Predicted $2p$ radioactive half-lives using different theoretical models and/or formulas for possible $2p$ radioactive nuclei whose $2p$ radioactivity is energetically allowed or observed but not yet quantified in NUBASE2016 \cite{50}.}
\label {fig 5}
\end{center}

\begin{multicols}{2}
\begin{center}
\includegraphics[width=9cm]{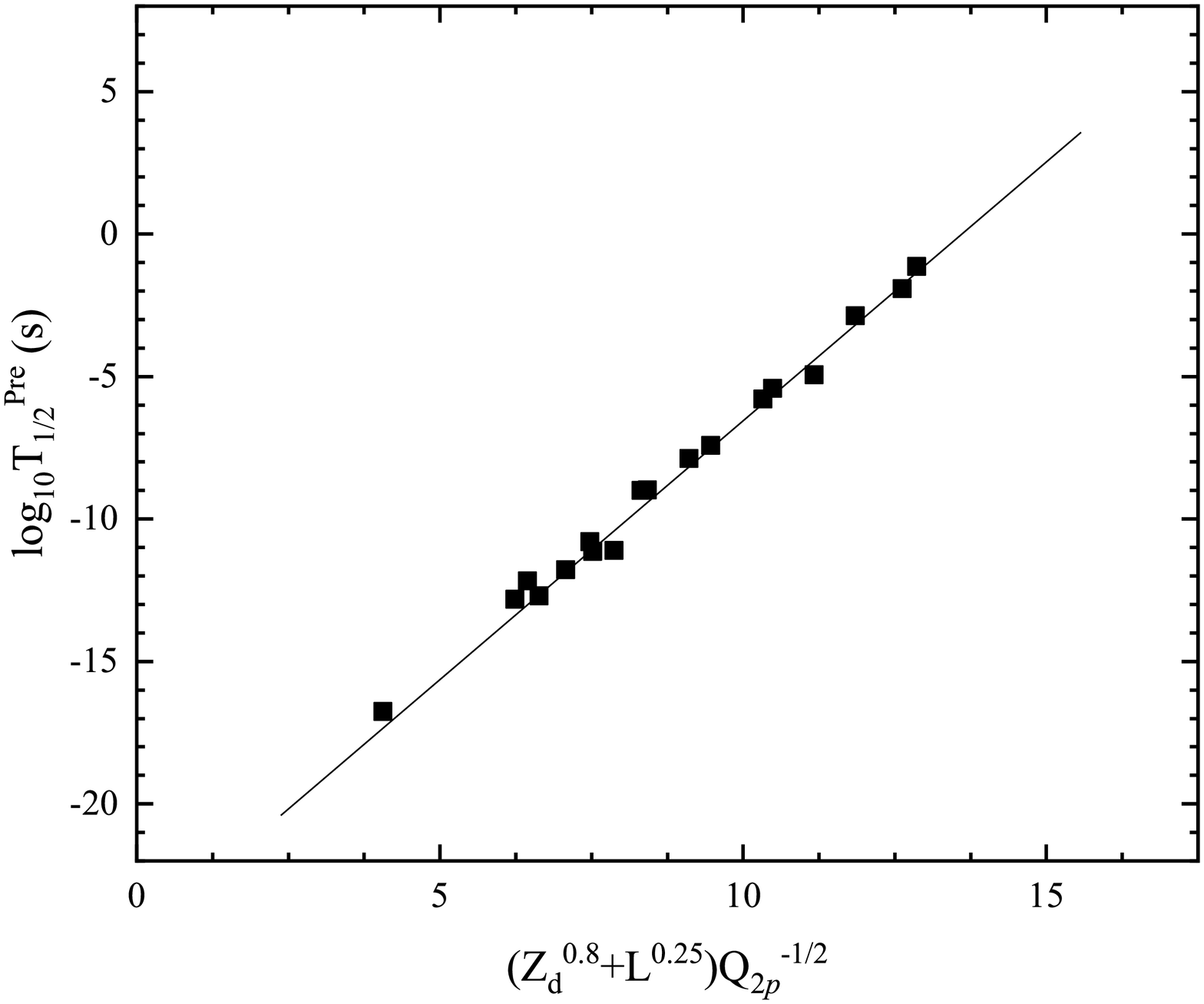}
\figcaption{\label{figure6} Predicted $2p$ radioactive half-lives $\log_{10}T_{1/2}^{\rm{Pre}}$ plotted against $(Z_d^{0.8} + l^{0.25})Q_{2p}^{-1/2}$, i.e., the new Geiger-Nuttall law for $2p$ radioactivity \cite{32}.}
\label {fig 6}
\end{center}
\end{multicols}

\begin{multicols}{2}
\section{Summary}
In summary, based on the WKB theory considering the electrostatic screening effect and using a Hulthen potential to replace the Coulomb potential, we systematically investigate the $2p$ radioactive half-lives of nuclei near or beyond the proton drip line. The screening parameter $a$ is obtained by fitting the experimental half-lives of true $2p$ radioactive nuclei according to the smallest standard deviation. The calculated results are found to be in agreement with the corresponding experimental data. In addition, we extend this model to predict the half-lives of possible $2p$ radioactive nuclei whose $2p$ radioactivity is energetically allowed or observed but not yet quantified in NUBASE2016. The predicted results are in agreement with the results calculated using the Gamow-like model, generalized liquid drop model, Sreeja formula, and Liu formula. Furthermore, there is an approximate linear trend between our predicted $2p$ radioactive half-lives $\log_{10}T_{1/2}^{Pre}$ and $(Z_d^{0.8} + l^{0.25})Q_{2p}^{-1/2}$, i.e., the new Geiger-Nuttall law for $2p$ radioactivity. It indicates that our predictions are reliable.
\end{multicols}
\vspace{-1mm}
\centerline{\rule{80mm}{0.1pt}}
\vspace{2mm}

\begin{multicols}{2}

\end{multicols}

\clearpage
\end{CJK*}
\end{document}